\documentstyle[11pt]{article}

\newcommand{\be}{\begin{equation}}
\newcommand{\ee}{\end{equation}}
\newcommand{\bea}{\begin{eqnarray}}
\newcommand{\eea}{\end{eqnarray}}
\newcommand{\non}{\nonumber}
\newcommand{\ra}{\rangle}

\begin{document}


\title{The $sl_2$ loop algebra symmetry of the twisted transfer matrix 
of the six vertex model  at roots of unity}

\author{Tetsuo Deguchi
\footnote{e-mail deguchi@phys.ocha.ac.jp}}

\date{}
\maketitle

\begin{center}
Department of Physics, Ochanomizu University,\\ 
2-1-1 Ohtsuka, Bunkyo-ku,Tokyo 112-8610, Japan
\end{center}

\begin{abstract}
We discuss a family of operators which commute or anti-commute 
with the twisted transfer matrix of 
the six-vertex model at $q$ being roots of unity: $q^{2N}=1$. 
The  operators commute with the Hamiltonian of the XXZ spin chain 
under the twisted boundary conditions,     
and they are valid also for the inhomogeneous case. 
For the case of the anti-periodic boundary conditions,   
we show explicitly that the operators generate 
the $sl_2$ loop algebra in the sector of the total spin 
operator $S^Z \equiv N/2 \, ({\rm mod} \, N)$. 
The infinite-dimensional symmetry 
leads to exponentially-large spectral degeneracies, as shown   
for the periodic boundary conditions \cite{DFM}.  
Furthermore, we derive explicitly the $sl_2$ loop algebra symmetry  
for the periodic XXZ spin chain 
with an odd number of sites 
in the sector  $S^Z \equiv N/2 \, ({\rm mod} \, N)$     
when $q$ is a primitive $N$ th root of unity with $N$ odd.  
Interestingly, in the case of $N=3$, 
 various conjectures of combinatorial formulas for the XXZ spin chain 
 with odd-sites have been given by Stroganov and others.  
%
%
  We also note a connection to the spectral degeneracies 
  of the eight-vertex model.     
\end{abstract}

\newpage 
\section{Introduction}

 Finite-size spectrum of the XXZ spin chain \cite{Bethe,Yang-Yang}
 under the twisted boundary conditions 
has attracted much interest recently and has been studied 
numerically or analytically such as by the Bethe ansatz 
\cite{Batchelor,SS,ACTWu,Fowler,Eckle,Itoyama}. 
 The XXZ Hamiltonian defined on a ring of $L$ sites is given by  
\be 
{\cal H}_{\rm XXZ} =  J \sum_{j=1}^{L} \left(\sigma_j^X \sigma_{j+1}^X +
 \sigma_j^Y \sigma_{j+1}^Y + \Delta \sigma_j^Z \sigma_{j+1}^Z  \right) \, . 
\label{hxxz}
\ee
Here $\sigma_j^{\alpha}$ for $\alpha=X,Y,Z$  denotes the Pauli matrix  
defined on the $j$th site. Under the periodic boundary conditions 
we have $\sigma_{L+1}^{\alpha} = \sigma_{1}^{\alpha}$ for any $\alpha$.
 We now introduce     
the twisted boundary conditions by  
\be 
\sigma_{L+1}^{\pm} = \exp(\pm i \Phi) \sigma_{1}^{\pm}  \, , 
 \qquad \sigma^{Z}_{L+1} = \sigma^{Z}_{1} . 
\ee
We call the parameter $\Phi$ the twist angle.  

\par  
There are several physical applications of the twisted boundary conditions. 
The twist angle corresponds to the magnetic flux threaded 
through the ring \cite{Byers-Yang}.  
Taking the variation of eigenvalues 
under changes of the flux \cite{Kohn}, 
the effective mass has been evaluated exactly 
for the many-body system of interacting fermions or bosons in one-dimension 
which is equivalent to the XXZ spin chain \cite{SS}.  
The finite-size spectrum of the XXZ spin chain under the twisted boundary 
conditions has also been studied  from the viewpoint of 
the finite-size analysis of conformal field theories \cite{Batchelor}. 
Associated with the variation of the ground-state energy, the flows of  
excited states have been numerically investigated with respect to  
 the twist angle   
\cite{Batchelor,SS,Fowler,Eckle}. 
 In the spectral flows, we find several level crossings 
 with large degeneracy. However, it  
  has not been discussed explicitly  
 what kind of symmetry operators  
 correspond to the spectral degeneracies.   
 The question should be interesting in particular  
 from the viewpoint of the violation of 
 the level non-crossing rule as discussed by 
 Heilmann and Lieb \cite{Heilmann-Lieb}.

 \par 
Recently, it is found that the  
symmetry of the XXZ spin chain 
becomes enhanced at roots of unity.  
The $sl_2$ loop algebra commutes with the XXZ Hamiltonian 
under the periodic boundary conditions,   
and the infinite-dimensional algebra leads to 
many spectral degeneracies whose dimensions 
increase exponentially with respect to the system size \cite{DFM}.    
Let us  introduce the  parameter $q$ through  
the XXZ anisotropy  $\Delta$ as    
\be 
\Delta= {\frac {q+q^{-1}} 2} 
\ee
It is  shown that 
the generators of the $sl_2$ loop algebra 
commute or anti-commute with the transfer matrix 
of the six-vertex model when $q^{2N}=1$, and 
 all the defining relations of the 
$sl_2$ loop algebra are explicitly derived 
 for the case in the sector $S^Z \equiv 0 \, ({\rm mod} \, N)$. 
 Here  $S^Z$ denotes 
the $Z$ component of the total spin operator. 
Several aspects of the $sl_2$ loop algebra symmetry of the 
XXZ spin chain has been studied 
\cite{FM,KM,Odyssey,Missing,FM-8V,Korff-aux,highest}. 
In particular, its connection to the spectral degeneracies 
of the transfer matrix of the eight-vertex model 
has been discussed \cite{Missing,FM-8V}. 
There are also some  
relevant papers \cite{BJ,Belavin}.

\par 
The purpose of the paper is  
to formulate  a family of operators  commuting 
with the XXZ Hamiltonian with the twisted boundary conditions. 
They may explain the level crossings observed in the spectral flows 
  with respect to the twist angle. Here   
we generalize the approach given in Ref. \cite{DFM},   
and give some extended results. 
As an illustration, we show that when $\Phi=\pi$ 
and  $q$ is a primitive $2N$th root of unity, 
the  operators commuting with the twisted XXZ Hamiltonian 
generate the $sl_2$ loop algebra for the sector 
 $S^Z = N/2 \, ({\rm mod} \, N)$.  
 Furthermore, in the sector $S^Z = N/2 \, ({\rm mod} \, N)$, 
 we explicitly show the defining relations of  the $sl_2$
loop algebra for the case of the periodic boundary 
conditions where $L$ is odd  and $q$ is a primitive 
$N$th root of unity with odd $N$. 
 When $N=3$, it is exactly the case in which  
 various combinatorial formulas were discussed recently 
  \cite{Odd,Razumov,BGN,GBNM}.

\par 
The content of the paper consists of the following: 
In \S 2 we review the $sl_2$ loop algebra symmetry of 
the periodic XXZ spin chain \cite{DFM}. In \S 3 we introduce the 
transfer matrix of the six-vertex model under the twisted 
boundary conditions. We also discuss some useful techniques such as 
the gauge transformation and the crossing symmetry. 
In \S 4 we show commutation relations of the twisted transfer matrix 
with some powers of the quantum group generators, which are 
fundamental in the paper. We also show those of the inhomogeneous case.     
 In \S 5 we discuss operators commuting with the twisted 
 transfer matrix when $q$ is a root of unity. 
We also discuss the special cases where we can explicitly check 
the defining relations of $sl_2$ loop algebra. 
Finally we note a connection to the eight-vertex model.

\section{The loop algebra symmetry 
of the periodic XXZ spin chain}

Let us review  the $sl_2$ loop algebra symmetry of the 
XXZ spin chain under the periodic boundary conditions. 
Let  us introduce the quantum group $U_q(sl_2)$. 
The generators $S^{\pm}$ and $S^Z$ satisfy the defining relations 
\be 
[S^{+}, S^{-}] = {\frac {q^{2 S^Z} - q^{-2 S^Z}} {q-q^{-1}} } \, , \quad  
[S^Z, S^{\pm}] =  \pm  S^{\pm} 
\ee
with the comultiplication $\Delta$ given by  
\be 
\Delta( S^{\pm}) = S^{\pm} \otimes q^{- S^Z} + q^{S^Z} \otimes S^{\pm} \, , 
\quad \Delta(S^Z) = S^Z \otimes I + I \otimes S^Z 
\ee
Here, the parameter $q$ is generic.   
In fact, we may consider $U_q(L(sl_2))$, i.e.,  
the $q$ analogue of the universal enveloping algebra of 
the $sl_2$ loop algebra.  For simplicity,  however, we only 
consider $U_q(sl_2)$ in the paper. 

\par 
Let $V$ denote a two-dimensional vector space over ${\bf C}$. 
 On the $L$th tensor product space $V^{\otimes L}$,   
 the generators $S^{\pm}$ and $S^{Z}$ are given by \cite{DFM}
\bea 
q^{S^Z} &= & q^{\sigma^Z/2} \otimes \cdots \otimes q^{\sigma^Z/2}  \non \\
S^{\pm} &= & \sum_{j=1}^{L} 
q^{\sigma^Z/2} \otimes \cdots \otimes q^{\sigma^Z/2} \otimes 
\sigma_j^{\pm} \otimes q^{-\sigma^Z/2} \otimes 
\cdots \otimes q^{-\sigma^Z/2}
= \sum_{j=1}^{L} S^{\pm}_j \quad . 
\eea
Here $S_j^{\pm}$ denotes the $j$th term with $\sigma_j^{\pm}$ 
in the sum (6).   
Considering the auto-morphism of $U_q(L(sl_2))$, 
we introduce the following operators: 
\be 
T^{\pm} = \sum_{j=1}^{L} 
q^{-\sigma^Z/2} \otimes \cdots \otimes q^{-\sigma^Z/2} \otimes 
\sigma_j^{\pm} \otimes q^{\sigma^Z/2} \otimes \cdots \otimes q^{\sigma^Z/2}   
= \sum_{j=1}^{L} T_j^{\pm} 
\ee
Let us denote by $S^{\pm(n)}$ and $T^{\pm(n)}$ the following operators:    
\be 
S^{\pm(n)} = \left( S^{\pm} \right)^n/[n]! \, , \quad 
T^{\pm(n)} = \left( T^{\pm} \right)^n/[n]!  
\ee
Here $n$ is a positive integer, and $[n]$ and $[n]!$ denote   
the $q$-integer $[n]=(q^n -q^{-n})/(q-q^{-1})$ 
and the $q$-factorial $[n]! = [n] [n-1] \cdots [1]$, respectively.  
Then,  we have   
\begin{eqnarray}
& & S^{\pm(n)} =  
\sum_{1 \le j_1 < \cdots < j_n \le L}
q^{{n \over 2 } \sigma^Z} \otimes \cdots 
\otimes q^{{n \over 2} \sigma^Z}
\otimes \sigma_{j_1}^{\pm} \otimes
q^{{(n-2) \over 2} \sigma^Z} \otimes  \cdots 
\nonumber \\
 & & \quad \otimes q^{{(n-2) \over 2} \sigma^Z} 
\otimes \sigma_{j_2}^{\pm} \otimes q^{{(n-4) \over 2} \sigma^Z} \otimes
\cdots
\otimes \sigma^{\pm}_{j_n} \otimes q^{-{n \over 2} \sigma^Z} \otimes \cdots
\otimes q^{-{n \over 2} \sigma^Z} \quad  
\label{sn} \\ 
& & T^{\pm(n)}  =  
\sum_{1 \le j_1 < \cdots < j_n \le L}
q^{- {n \over 2 } \sigma^Z} \otimes \cdots \otimes q^{- {n \over 2} \sigma^Z}
\otimes \sigma_{j_1}^{\pm} \otimes
q^{- {(n-2) \over 2} \sigma^Z} \otimes  \cdots 
\nonumber \\
 & & \quad \otimes q^{ - {(n-2) \over 2}\sigma^Z} 
\otimes \sigma_{j_2}^{\pm} \otimes q^{- {(n-4) \over 2} \sigma^Z} \otimes
\cdots
\otimes \sigma^{\pm}_{j_n} \otimes q^{{n \over 2} \sigma^Z} \otimes \cdots
\otimes q^{{n \over 2} \sigma^Z} \quad . 
\label{tn}
\end{eqnarray}

\par 
Let the symbol $\tau_{6V}(v)$ denote the transfer matrix 
of the six-vertex model. 
We now take the parameter $q$ a root of unity.  
We consider the limit of sending $q$ to a root of unity:  $q^{2N}=1$.     
Then we can show the (anti) commutation relations \cite{DFM} 
in the sector of $S^Z \equiv 0$ (mod $N$) 
\begin{equation}
S^{\pm (N)} \tau_{6V}(v)=q^N \tau_{6V}(v) S^{\pm (N)}, \qquad 
T^{\pm (N)} \tau_{6V}(v)=q^N \tau_{6V}(v) T^{\pm (N)} 
\end{equation} 
Since the XXZ Hamiltonian $H_{XXZ}$ is given by 
the logarithmic derivative of the transfer matrix,     
we have  in the sector $S^Z \equiv 0$ (mod $N$)  
\begin{eqnarray}
{[}S^{\pm(N)},H_{XXZ} {]}={[}T^{\pm(N)},H_{XXZ} {]}=0.
\label{sthcomm}
\end{eqnarray}

\par 
Let us  now consider the algebra generated by 
 the operators \cite{DFM}. 
When $q$ is a primitive $2N$th root of unity with $N$ even, 
or a primitive $N$th root of unity with $N$ odd, 
we consider the following  identification \cite{DFM}:    
\be
E_0^{+}=S^{+(N)}, \, E_0^{-}=S^{-(N)}, \, E_1^{+}=T^{-(N)}, \, 
E_1^{-}=T^{+(N)}, \, H_0=-H_1= {\frac 2 N}  S^Z  .    
\ee
Using the auto-morphism 
\be 
\theta(E_0^{\pm})=E_1^{\pm} \,  , \quad \theta(H_0)=  H_1 
\ee
we may take the following identification:   
\begin{equation}
E_0^{+} = T^{-(N)}, \, E_0^{-}=T^{+(N)}, \, E_1^{+} = S^{+(N)}, \, 
E_1^{-} = S^{-(N)} , \, -H_0=H_1= {\frac 2 N}  S^Z \, .    
\label{id2}
\end{equation}
When $q$ is a primitive $2N$th root of unity with $N$ odd,    
we may put as follows 
\begin{equation}
E_0= i T^{-(N)}, \,  E_1= i S^{+(N)}, \, F_0= i T^{+(N)}, \,
F_1= i S^{-(N)},  
 \, -H_0=H_1=  {\frac 2 N}  S^Z  .   
\label{imaginary}
\end{equation}
It is shown in Ref. \cite{DFM} 
that the operators $E_j^{\pm}, H_j$ for $j=0,1$,  
satisfy the defining relations  of the algebra $U(L(sl_2))$.
Explicitly they are given by the following:  
\bea 
& &  H_0 + H_1  = 0  , \, 
{\rm [} H_i, E_j {\rm ]}  =  a_{ij} E_j  ,  \, 
 {\rm [} H_i, F_j {\rm ]}   = - a_{ij} F_j  ,   \quad (i,j = 0, 1) 
 \label{Cartan} \\ 
& &  {\rm [} E_i, F_j {\rm ]}  =  \delta_{ij} H_{j} \, , 
  \qquad (i,j = 0, 1) \label{EF} \\
& & {\rm [} E_i,  {\rm [} E_i, {\rm [} E_i, 
 E_j {\rm ]}   {\rm ]}   {\rm ]} = 0 , \,  
{\rm [} F_i,  {\rm [} F_i, {\rm [} F_i,  F_j {\rm ]}   {\rm ]} =0   
\, (i,j=0,1, \, i \ne j) 
\label{Serre} 
\eea 
Here, the Cartan matrix $(a_{ij})$ of $A_1^{(1)}$ is defined  by 
\be 
\left( 
\begin{array}{cc}  
a_{00} & a_{01} \\
a_{10} & a_{11} 
\end{array} 
\right) 
 = \left( 
\begin{array}{cc} 
2 & -2 \\
-2 & 2 
\end{array}
\right)
\ee

\par 
The relations (\ref{Cartan}) hold for generic $q$, 
while the higher Serre relations (\ref{Serre}) hold if $q$ is a 
primitive $2N$th root of unity 
or a primitive $N$th root of unity with $N$ odd \cite{DFM,Lusztig}.  
The relation (\ref{EF}) holds 
if $q$ is a primitive $2N$th root of unity with $N$ even 
or a primitive $N$th root of unity with $N$ odd.   
 When $q$ is a primitive $2N$th root of unity with $N$ odd, 
we take the identification (\ref{imaginary}) with the imaginary factors.

\section{Twisted transfer matrix of the six-vertex model }

\subsection{Boltzmann weights}
Let us consider the configuration around the vertex as shown in Fig. 1. 
Variables $a, b, c$ and $d$ are defined on the edges 
at the vertex, and they take values 1 or 2. 
The value 1 corresponds to a polarization vector that  
 is in the  upward or rightward directions, 
and the value 2 to a  polarization vector 
in the downward or leftward direction.    
We assign the Boltzmann weight $X^{ac}_{bd}(u)$ 
to the configuration in Fig. 1.  The weight vanishes unless 
$a+c=b+d$, which we call the `ice rule' or the `charge conservation'.  
All the nonzero Boltzmann weights are given by  
\bea
X^{11}_{11}(u) & = & X^{22}_{22}(u) = \sinh (2 \eta + u) \qquad 
X^{12}_{21}(u)  =  X^{21}_{12}(u) = \sinh u \non \\
& & X^{12}_{12}(u)  = X^{21}_{21}(u) = \sinh 2 \eta 
\label{sbw} 
\eea
Here $q=\exp 2 \eta$, and $u$ is the spectral parameter. 

\par 
We define operators $X_j(u)$s  for $j=0, 1, \ldots, L-1$ by 
\begin{equation} 
{X}_j(u) = 
\sum_{a,b,c,d =1,2 } {X}^{ ac}_{ bd}(u) I_0 \otimes 
I_1 \otimes \cdots \otimes I_{j-1} \otimes E^{ab}_{j} \otimes E^{cd}_{j+1}  
\otimes I_{j+2} \otimes \cdots \otimes I_L  
\label{tensor} 
\end{equation}  
where  $I$ denotes the identity matrix and $E^{ab}$ denotes the matrix  
\begin{equation} 
\left( E^{ab} \right)_{c, d} 
= \delta_{a,c } \, \delta_{b, d} \quad {\rm for} \quad  c , d = 1, 2 \, .  
\end{equation}

\par 
It is easy to show that the operators 
$X_j(u)$s constructed from the Boltzmann weights (\ref{sbw}) 
satisfy the Yang-Baxter equation in the following:   
\be 
X_j(u)X_{j+1}(u+v)X_j(v) = X_{j+1}(v)X_j(u+v)X_{j+1}(u) 
\label{YBE-X}
\ee
In terms of the Boltzmann weights the operator relation (\ref{YBE-X}) 
is written as follows 
\be 
\sum_{\alpha, \beta, \gamma} X^{a_1 a_2}_{\alpha \gamma}(u) 
X^{\gamma a_3}_{\beta b_3}(u+v) X^{\alpha \beta}_{b_1 b_2}(v) 
= 
\sum_{\alpha, \beta, \gamma} X^{a_2 a_3}_{\beta \alpha}(v) 
X^{a_1 \beta}_{b_1 \gamma}(u+v) X^{\gamma \alpha}_{b_2 b_3}(u) 
\label{YBE-w}
\ee

\begin{figure}[htb] 
\begin{center}
\setlength{\unitlength}{0.8mm}
\begin{picture}(160,40)(0,0)
\put(100,20){\thicklines\line(-1,0){20}}
\put(80,20){\thicklines\line(-1,0){20}}
\put(80,20){\thicklines\line(0,1){20}}
\put(80,0){\thicklines\line(0,1){20}}
\put(60,20){\makebox(5,5){$b$}}
\put(80,0){\makebox(5,5){$d$}}
\put(80,35){\makebox(5,5){$a$}}
\put(95,20){\makebox(5,5){$c$}}
\end{picture}
\end{center}
\caption{Vertex configuration for the Boltzmann weight $X^{ac}_{bd}(u)$. 
The spectral parameter $u$ corresponds to the angle 
between the two lines $b$ to $c$ 
and $d$ to $a$, which is important to the Yang-Baxter equation (\ref{YBE-w}).} 
\label{fig1}
\end{figure}
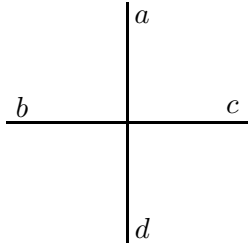 

\subsection{The six-vertex transfer matrix with 
the twisted boundary conditions }

\par  
Recall that the operators $X_j(u)$s are acting on the tensor product 
$V^{\otimes (L+1)}= V_0 \otimes V_1 \otimes \cdots \otimes V_L$. 
We now define the twisted transfer matrix of the six-vertex model by     
\be 
\tau(u ; \phi)= {\rm tr}_0 \left( q^{\phi \sigma_0^Z} 
X_{L-1}(u) \cdots X_1(u) X_0(u) \right)
\label{twist}
\ee
where the symbol ${\rm tr}_0$ denotes the trace over the 0th space $V_0$.

\par 
The logarithmic derivative of the twisted transfer matrix 
leads to  the XXZ Hamiltonian ${\cal H}_{XXZ}(\phi)$ 
under the twisted boundary conditions  
\bea 
& &  {\sinh 2 \eta}  \times  \, {\frac d {du} } \log \tau(u; \phi)|_{u=0}  
=  \sum_{j=1}^{L-1} 
\left( 2 \sigma_{j}^{+} \sigma_{j+1}^{-} +  
2 \sigma_{j}^{-} \sigma_{j+1}^{+} + \cosh 2 \eta  
\sigma_{j}^{Z} \sigma_{j+1}^{Z} \right) 
\non \\
&  & \qquad + \,  
  q^{- 2\phi} \, 2 \sigma_{L}^{+} \sigma_{1}^{-} +  
q^{2 \phi} \, 2 \sigma_{L}^{-} \sigma_{1}^{+} + \cosh 2 \eta  
\sigma_{L}^{Z} \sigma_{1}^{Z}  +  L \cosh 2 \eta  \non \\
&  & \qquad = \,  
{\cal H}_{XXZ}(\phi)/J + L \Delta \, , 
\label{twistedXXZ}
\eea
where $\Delta = \cosh 2 \eta$ and 
the twist angle $\Phi$ is related to $\phi$ through the relation 
\be 
q^{2 \phi} = \exp(i \Phi)
\ee

\par 
We should  note that the twisted transfer matrix (\ref{twist}) is 
different from that of Ref. \cite{DFM}:  
 the logarithmic derivative of the transfer matrix 
$T^{DFM}(v; \phi)$ of Ref. \cite{DFM} leads to the XXZ Hamiltonian 
under the periodic boundary conditions 
 for any $\phi$, while that of 
   (\ref{twist}) gives the twisted XXZ Hamiltonian (\ref{twistedXXZ}).

\par 
 Let the symbol $\Pi^{(12 \cdots L)}$ denote the shift operator defined by  
\be 
\Pi^{(12\cdots L)} e_1 \otimes \cdots \otimes e_L = 
e_L \otimes e_1 \otimes \cdots \otimes e_{L-1} 
\ee
Then,  the twisted transfer matrix $\tau(u; \phi)$
gives the twisted shift operator: 
$\tau(0; \phi ) = \sinh^L 2\eta \, \Pi(\phi)$,  
where  $\Pi(\phi)$ is defined by 
\be  
\Pi(\phi) = q^{\phi \sigma_1^Z} \Pi^{(12\cdots L)} 
=  \Pi^{(12\cdots L)} q^{\phi \sigma_L^Z} 
\ee

\subsection{Gauge transformations and the crossing symmetry}

We introduce the following transformation on the 
Boltzmann weights of the six-vertex model\cite{AW}:  
\be 
X^{ac}_{bd}(u) \rightarrow {\tilde X}^{ac}_{bd}(u) 
= \epsilon^{a+b} \, \exp \left(\kappa (a+b-c-d)u/2 \right) X^{ac}_{bd}(u)
\label{gauge}
\ee
Here $\kappa$ is arbitrary and $\epsilon= \pm 1$.  
We can show that 
the transformed weights ${\tilde X}^{ac}_{bd}(u)$s satisfy the 
Yang-Baxter equation (\ref{YBE-w}) if  $X^{ac}_{bd}(u)$s satisfy it.   
We call  (\ref{gauge}) a gauge transformation.  
Let ${\tilde X}_j(u)$ denote the operator defined by (\ref{tensor}) 
with the $X^{ac}_{bd}(u)$ replaced with ${\tilde X}^{ac}_{bd}(u)$. 
Then, ${\tilde X}_j(u)$s also satisfy the Yang-Baxter equation  
(\ref{YBE-X}).

\par 
Let us discuss the gauge transformation (\ref{gauge})  
with $\epsilon=1$. 
It has two important properties. 
 First,  the transfer matrix $\tau(u; \phi)$ is invariant under the   
the gauge transformation due to the charge conservation. 
 Second, when $\kappa=\pm 1$, 
 ${\tilde X}_j(u)$  can be expressed 
 in terms of the generator of the Temperley-Lieb algebra \cite{AW}. 
Let ${X}^{\pm ac}_{~~bd}(u)$ denote  
the transformed weight ${\tilde X}^{ac}_{bd}(u)$ with $\kappa= \pm 1$, 
respectively. In terms of the Boltzmann weights, we have the 
following decomposition \cite{AW}:      
\be 
{X}^{\pm ac}_{~~bd}(u) 
 =  \sinh(u+ 2 \eta) \delta_{a,b} \delta_{c,d} 
+  \sinh u \, r_a^{\pm} r_b^{\pm}  \, 
\delta_{a, {\bar c}} 
\delta_{b, {\bar d}} \,  
\label{Xdc} 
\ee 
Here ${\bar a}$ denotes the conjugate of $a$, 
which is defined by  ${\bar 1}= 2$ and 
${\bar 2} = 1$. 
The quantities $r_j^{\pm}$ $(j=1,2)$ are defined by \cite{DWA}
\be 
r_1^{\pm} =  i \exp( \mp \eta) \, , 
\quad r_2^{\pm} = - i \exp( \pm \eta) 
\ee

\par 
The weights ${X}^{\pm \, a,c}_{~~\, b,d}(u)$s  
have the following symmetry \cite{DWA}:  
 \begin{equation}
 {X}^{\pm \, a,c}_{~~\, b,d}(u) 
 = - r_b^{\pm} r_{\bar c}^{\pm} 
 {X}^{\pm \, {\bar b}, a}_{~~\, d, {\bar c} }(-2 \eta-u)  
= - r_a^{\pm} r_{\bar d}^{\pm}  
 {X}^{\pm \, c, {\bar d}}_{~~\, {\bar a}, b }(-2 \eta-u)  
 \label{crossing} 
 \end{equation}
We call it the crossing symmetry.

\subsection{The Temperley-Lieb decomposition of $X_j(u)$s}

Let us introduce the following operators 
\begin{equation} 
{U}_j^{\pm} = 
\sum_{a,b,c,d =1,2 } r_a^{\pm} r_b^{\pm}  \, 
\delta_{a, {\bar c}} \delta_{b, {\bar d}} 
 I_0 \otimes I_1 \otimes \cdots \otimes 
 I_{j-1} \otimes E^{ab}_{j} \otimes E^{cd}_{j+1}  
\otimes I_{j+2} \otimes \cdots \otimes I_L  
\label{U} 
\end{equation}  
The $U_j^{\pm}$s satisfy the defining relations 
of the Temperley-Lieb algebra \cite{TL,Inversion}:
\be 
(U_j^{\pm})^2  =   Q^{1/2} U^{\pm}_j, \quad 
U_j^{\pm} U^{\pm}_{j+1} U^{\pm}_j  = U^{\pm}_j,   \quad 
U^{\pm}_{j+1} U^{\pm}_{j} U^{\pm}_{j+1} = U^{\pm}_{j+1},  
\ee
for $j=1,2, \ldots, L-1$.  
Here $Q^{1/2}$ is given by $Q^{1/2}=- (q + q^{-1})$.    
We call $U_j^{\pm}$s the Temperley-Lieb operators. 

\par In terms of the Temperley-Lieb operators, 
the operator $X_j^{\pm}(u)$ can be expressed as follows 
\cite{Inversion}
\be 
X_j^{\pm}(u) = \sinh(u+ 2\eta) \, I + \sinh u \, U_j^{\pm}
\label{TLdc}
\ee
The decomposition (\ref{TLdc}) corresponds to  (\ref{Xdc}) 
for the Boltzmann weights. 

\par 
The Temperley-Lieb operators $U_j^{+}$s 
commute with the generators of $U_q(sl_2)$. We can show    
$ {[} S^{\pm}, U_j^{+} {]} = 0$ for $j=0, 1, \ldots, L-1$.  
 It can be important, since  the XXZ Hamiltonian can be expressed  
in terms of $U_j^{+}$s and a boundary term \cite{PS}. 

\par 
It is  known that the periodic XXZ spin chain does not 
commute with $U_q(sl_2)$. In fact, the generators 
$S^{\pm}$ do not commute 
with the periodic XXZ Hamiltonian, 
since they are not compatible with the periodic boundary conditions. 
 When $q$ is a root of unity, however, some powers of 
$S^{\pm}$ can commute with it \cite{PS}.  
As a matter of fact, the observation can be developed 
much further  \cite{DFM}. 
We first note that $U_j^{+}$s commute with $S^{\pm}$ while 
 they do not with $T^{\pm}$, and    
$U_j^{-}$s commute with $T^{\pm}$ while 
 they do not with $S^{\pm}$.  when $q^{2N}=1$, we can show that 
 the transfer matrix commutes with $S^{\pm (N)}$ and $T^{\pm (N)}$ 
 simultaneously, and they generate the $sl_2$ loop algebra \cite{DFM}.

\section{Relations of the transfer matrix for $q$ generic}

\subsection{Decomposition of the transfer matrix}
 
Let us consider the gauge transformations (\ref{gauge}) with 
$\kappa = \pm 1$ and $\epsilon =1$.  
Due to the charge conservation, they do not change any of  
the matrix elements 
of the transfer matrix (\ref{twist}). Thus, we have  
\be 
\tau(u; \phi) =   {\rm tr}_0 
\left(q^{\phi \sigma_0^Z} X_{L-1}^{\pm}(u) \cdots 
X_1^{\pm}(u) X_0^{\pm}(u) \right) 
\label{transformed}
\ee
We denote by $\tau^{\pm}(u; \phi)$ the right hand side of
(\ref{transformed}). 
Putting  the decomposition (\ref{TLdc}) for $X_0^{\pm}(u)$ 
into $\tau^{\pm}(u; \phi)$,  
and making use of the crossing symmetry 
(\ref{crossing}),  we can show the following: 
\be 
\tau(u; \phi)   = \Pi(\phi) X_{LL}^{\pm}(u) 
+ X_{RR}^{\pm}(u) \Pi(\phi)^{-1} \, , 
\label{dc} 
\ee
where the symbols $X_{LL}^{\pm}(u)$ and $X_{RR}^{\pm}(u)$ are given by  
\bea 
X_{LL}^{\pm}(u) & = & \sinh( u + 2 \eta) \, 
X_{L-1}^{\pm}(u) \cdots X_{2}^{\pm}(u) X_{1}^{\pm}(u)
\non \\
X_{RR}^{\pm}(u) & = & (-1)^L \sinh (u) \, 
 X_1^{\pm}(-2 \eta -u) X_2^{\pm}(-2 \eta -u) \cdots 
X_{L-1}^{\pm}(-2 \eta -u) \non \\
\eea
We remark that $X_j^{+}(u)$s commute with $S^{\pm}$ while  
 $X_j^{-}(u)$s commute with $T^{\pm}$. 
When $\phi=0$, (\ref{dc}) 
is reduced to that of the periodic one \cite{DFM}.

\subsection{Transformations of the twisted shift operator}

We can show the following relations 
of $(S^{\pm})^n$ and $(T^{\pm})^{n}$ 
for  generic $q$:    
\bea 
& & \Pi(\phi) \left( S^{\pm} \right)^n \Pi(\phi)^{-1} 
 = q^{-n\sigma_1^Z}  \Big\{ (S^{\pm})^n + 
q^{\mp(n-1)}[n] (S^{\pm})^{n-1}S_1^{\pm} 
\left( q^{2(S^Z \pm n  \pm \phi)} -1 \right) \Big\}   \non \\  
& & \Pi(\phi)^{-1} \left( S^{\pm} \right)^n \Pi(\phi)  
  =   q^{n \sigma_L^Z}  \Big\{ (S^{\pm})^n + 
q^{\pm (n-1)}[n] (S^{\pm})^{n-1} S_L^{\pm} 
\left( q^{-2(S^Z \pm n  \pm \phi)} -1 \right) \Big\} \non  \\
& & \Pi(\phi) \left( T^{\pm} \right)^n  \Pi(\phi)^{-1} 
 =  q^{n \sigma_1^Z}  \Big\{ (T^{\pm})^n + 
q^{\pm(n-1)}[n] (T^{\pm})^{n-1} T_1^{\pm} 
\left( q^{-2(S^Z \pm n  \mp \phi)} -1 \right) \Big\}  \non \\   
& & \Pi(\phi)^{-1} \left( T^{\pm} \right)^n \Pi(\phi)  
 =  q^{-n \sigma_L^Z}  \Big\{ (T^{\pm})^n + 
q^{\mp (n-1)}[n] (T^{\pm})^{n-1} T_L^{\pm} 
\left( q^{2(S^Z \pm n  \mp \phi)} -1 \right) \Big\}  \non \\
 \label{PPP}  
\eea


\par 
Let us briefly discuss the derivation. 
We first consider the following:  
\bea 
 \Pi(\phi) \left( S^{\pm} \right)^n  \Pi(\phi)^{-1} 
& = & \left( \Pi(\phi) S^{\pm} \Pi(\phi)^{-1} \right)^n \non \\
& = & \left\{(S^{\pm}-S_1^{\pm}) q^{-\sigma_1^Z} 
+ q^{2 \phi} S_1^{\pm} q^{2S^Z-\sigma_1^Z} \right\}^n       
\eea
Here we denote by  $A$ and $q^{\pm 2\phi} B$ 
 the first and second terms of the 
right hand side, respectively. Then we can show the following:   
\be 
\left(A+ q^{\pm 2 \phi} B \right)^n = 
A^n + q^{\pm 2 \phi} \sum_{j=0}^{n-1} A^{n-1-j} B A^{j} 
\ee
Thus, we can derive the expression (\ref{PPP}) through the following:      
 \bea 
 A^{n} & = & q^{-n \sigma_1^Z} \left(S^{\pm} - S_1^{\pm} \right)^{n} 
 =  q^{-n \sigma_1^Z} \left( (S^{\pm})^n 
 - q^{\mp(n-1)}[n](S^{\pm})^{n-1} S_1^{\pm} \right) \non \\  
& & A^{n-1-j} B A^{j}  =  q^{\pm 2 (j+1)} q^{-n \sigma_1^Z} 
(S^{\pm})^{n-1} S_1^{\pm} q^{2 S^Z}
\eea

\subsection{Relations of the twisted transfer matrix }

Using the relations (\ref{PPP}), we can derive   
the following relations of the twisted transfer matrix 
for generic $q$: 
\bea 
& & (S^{\pm})^n \, \tau(u; \phi) =  \tau(u; \phi + n) \, (S^{\pm})^n \non \\
& & \qquad + q^{\pm (n-1)}[n] \Pi(\phi+n) (S^{\pm})^{n-1} S_L^{\pm} 
\left(q^{-2(S^Z \pm n \pm \phi)} -1 \right) \, X_{LL}^{+}(u)        
\non \\
&  & \qquad + q^{\mp (n-1)}[n] X_{RR}^{+}(u) \, \Pi(\phi+n)^{-1} 
 (S^{\pm})^{n-1} S_1^{\pm} 
\left(q^{2(S^Z \pm n \pm  \phi)} -1 \right) \, \non \\ 
& & 
(T^{\pm})^n \, \tau(u; \phi)  =  \tau(u; \phi-n) \, (T^{\pm})^n \non \\
& & \qquad + q^{\mp (n-1)}[n] \Pi(\phi-n) (T^{\pm})^{n-1} T_L^{\pm} 
\left(q^{2(S^Z \pm n  \mp  \phi)} -1 \right) \, X_{LL}^{-}(u)        
\non \\
&  & \qquad + q^{\pm (n-1)}[n] X_{RR}^{-}(u) \, \Pi(\phi-n)^{-1} 
(T^{\pm})^{n-1} T_1^{\pm} 
\left(q^{- 2(S^Z \pm n \mp  \phi)} -1 \right) \non \\        
& & \label{STtr} 
\eea
 Here we note that $S^{\pm}$ commute with $X^{+}_{LL}(u)$ and $X^{+}_{RR}(u)$,   and also that $T^{\pm}$ commute with $X^{-}_{LL}(u)$ and $X^{-}_{RR}(u)$.    
Some different forms of the commutation relations are given in Appendix A.

\par 
We remark that some relations of  $S^{\pm (N)}$ 
with the transfer matrix have been investigated \cite{BG}.

\subsection{Inhomogeneous case of the twisted transfer matrix}

We define the inhomogeneous transfer matrix of the six-vertex model 
under the twisted boundary conditions by  
\be 
\tau(u ; \phi; \{ w_j \})= {\rm tr}_0 
\left( q^{\phi \sigma_0^Z} 
X_{L-1}(u-w_{L-1}) \cdots X_1(u-w_1) X_0(u-w_0) \right)
\label{inhomo}
\ee
Here $w_j$s are called inhomogeneous parameters. 
The twisted transfer matrix (\ref{inhomo}) is called inhomogeneous.  
Under the gauge transformation (\ref{gauge}) with $\epsilon=1$,   
${\tau}(u; \phi; \{ w_j \})$ is mapped to 
 ${\tilde \tau}(u; \phi; \{ w_j \})$ as follows  
\be 
\tau(u; \phi; \{ w_j \} ) 
= V(\kappa) \, {\tilde \tau}(u; \phi; \{ w_j \} ) \, V(\kappa)^{-1}.  
\ee
where $V(\kappa)$ is given by the diagonal matrix  
\be 
\left( V(\kappa) \right)^{a_1, a_2, \cdots, a_L}_{b_1, b_2, \cdots, b_L}
= \exp\left(\kappa \sum_{j=1}^{L} w_j a_j \right) \delta_{a_1, b_1} 
\delta_{a_2, b_2} \cdots \delta_{a_L, b_L}
\ee
Let $\tau^{\pm}(u; \phi; \{ w_j \})$ denote the transformed transfer matrices  
${\tilde \tau}(u; \phi; \{ w_j \})$ with $\kappa=\pm 1$, 
respectively. We write $V(\pm 1)$ by ${V}^{\pm}$.
 We thus have 
\be 
\tau(u; \phi; \{ w_j \}) 
= {V}^{\pm} \,  \tau^{\pm}(u; \phi; \{ w_j \})  \,  {V}^{\mp} 
\label{VTV}
\ee
 We therefore define ${\tilde S}^{\pm}$ and ${\tilde T}^{\pm}$ by 
 \be 
 {\tilde S}^{\pm} = V^{+} \, S^{\pm} \, V^{-} \qquad 
 {\tilde T}^{\pm} = V^{-} \, T^{\pm} \, V^{+} . 
 \ee

\par 
The transfer matrices  $\tau^{\pm}(u; \phi; \{ w_j \})$ can be decomposed as  
\be 
\tau^{\pm}(u; \phi; \{w_j \}) = \Pi(\phi) \,  X_{LL}^{\pm}(u; \{ w_j \}) 
+ X_{RR}^{\pm}(u; \{ w_j \}) \, \Pi(\phi)^{-1} \, , 
\label{dc-inhomo} 
\ee
where $X_{LL}^{\pm}(u; \{ w_j \})$ and  $X_{RR}^{\pm}(u; \{ w_j \})$ 
are given by  
\bea 
& & X_{LL}^{\pm}(u; \{ w_j \}) = \sinh( u + 2 \eta- w_0) \,  
X_{L-1}^{\pm}(u-w_{L-1}) \cdots 
X_{1}^{\pm}(u-w_1) 
\non \\
& & X_{RR}^{\pm}(u; \{ w_j \}) = (-1)^L \sinh (u-w_0) \,   
 X_1^{\pm}(-2 \eta -u + w_1) \non \\ 
 & & \qquad \quad \times X_2^{\pm}(-2 \eta -u + w_2) \cdots 
X_{L-1}^{\pm}(-2 \eta -u + w_{L-1}) 
\eea
 In the same way as (\ref{STtr}) we can show  
 \bea 
& & (S^{\pm})^n \, \tau^{+}(u; \phi; \{ w_j \}) 
= \tau^{+}(u; \phi + n;  \{ w_j \}) \, (S^{\pm})^n \non \\
&  & \qquad + q^{\pm (n-1)}[n] \Pi(\phi+n) (S^{\pm})^{n-1} S_L^{\pm} 
\left(q^{-2(S^Z\pm n \pm \phi)} -1 \right) X_{LL}^{+}(u;  \{ w_j \})        
\non \\
& & \qquad +  q^{\mp (n-1)}[n] X_{RR}^{+}(u;  \{ w_j \}) \Pi(\phi+n)^{-1} 
 (S^{\pm})^{n-1} S_1^{\pm} 
\left(q^{2(S^Z \pm n \pm  \phi)} -1 \right) \non \\ 
& & (T^{\pm})^n \tau^{-}(u; \phi;  \{ w_j \}) 
 =  \tau^{-}(u; \phi-n;  \{ w_j \}) (T^{\pm})^n \non \\
& & \qquad + q^{\mp (n-1)}[n] \Pi(\phi-n) (T^{\pm})^{n-1} T_L^{\pm} 
\left(q^{2(S^Z \pm n  \mp  \phi)} -1 \right)  
X_{LL}^{-}(u; \{ w_j \})        
\non \\
& & \qquad + q^{\pm (n-1)}[n] X_{RR}^{-}(u; \{ w_j \}) \Pi(\phi-n)^{-1} 
(T^{\pm})^{n-1} T_1^{\pm} 
\left(q^{- 2(S^Z \pm n \mp  \phi)} -1 \right)  \non \\      
& & \label{tr-inhomo} 
\eea 
 Through the relation (\ref{VTV}) we can derive  commutation  
 or anti-commutation relations 
 for  the inhomogeneous twisted transfer matrix  $\tau(u; \phi; \{ w_j \})$. 
Let $|k \ra \ra$ denote such a vector with $S^Z=k$. Then we have 
\bea 
({\tilde S}^{\pm})^n \, \tau(u; \phi; \{ w_j \}) \, | k \ra \ra
& = & \tau(u; \phi+n; \{ w_j \}) \, ({\tilde S}^{\pm})^n \, 
 | k \ra \ra \, , 
\, {\rm when} \, q^{2(k \pm n \pm \phi)} = 1 \non \\ 
({\tilde T}^{\pm})^n \, \tau(u; \phi; \{ w_j \}) \, | k \ra \ra 
& = & \tau(u; \phi-n; \{ w_j \}) \, ({\tilde T}^{\pm})^n \, 
 | k \ra \ra \, , 
\, {\rm when} \, q^{2(k \pm n \mp \phi)} = 1  \non \\
& & \label{inhomo-main}
\eea
We note that for the case of $\phi=0$, the inhomogeneous 
result was addressed at the end of Ref. \cite{DFM}.

\section{The loop algebra symmetry}

\subsection{Operators commuting with the twisted transfer matrix}

Let us  assume that $q$ is a root of unity: $q^{2N}=1$. 
We denote by $| k \ra$ such a vector that has a fixed $S^Z$ value 
and it is equivalent to $k$ mod $N$: $S^Z \equiv k$ (mod $N$).    
Here $k$ is an integer or a half-integer.  
 Let $m$ and $n$ be two non-negative integers such that 
 $m \equiv n$ (mod $N$). 
 Then we have  
\bea
\left( S^{\pm} \right)^{m} \left( T^{\mp} \right)^{n} \tau(u; \phi) 
|\pm n \mp \phi \ra & = & \tau(u; \phi+m-n) 
\left( S^{\pm} \right)^{m} \left( T^{\mp} \right)^{n} | \pm n \mp \phi \ra 
\non \\
\left(T^{\pm}\right)^{m} \left(S^{\mp}\right)^{n} \tau(u; \phi) 
| \pm n \pm \phi \ra & = & \tau(u; \phi-m+n) 
\left(T^{\pm}\right)^{m} \left(S^{\mp}\right)^{n} |\pm n \pm \phi \ra 
\non \\
& & \label{ST}
\eea
The relations (\ref{ST}) are derived from (\ref{STtr}).  
We note that when $m=n$  the expressions (\ref{ST}) 
are also valid for the case of $q$ generic.

\par 
Let us denote by $\theta(n)$ the least non-negative integer 
which is equivalent to $n$ mod $N$.  
In the sector  $S^Z \equiv \ell$ (mod $N$) we have commutation 
relations $X \tau(u, p) = \tau(u, p) X$, where $X$s are given by     
\be 
(S^{\pm})^{\theta(\pm \ell+p)} (T^{\mp})^{\theta(\pm \ell+p)} \,,  \quad 
(T^{\pm})^{\theta(\pm \ell - p)} (S^{\mp})^{\theta(\pm \ell - p)}\, , 
\label{op1}
\ee
and commutation or anti-commutation relations 
$X \tau(u, p)$ = $q^{N} \tau(u, p) X$, where $X$s are given by 
\bea 
& & (S^{\pm})^{\theta(\pm \ell+p)+N} (T^{\mp})^{\theta(\pm \ell+p)} \, , 
\qquad 
(S^{\pm})^{\theta(\pm \ell+p)} (T^{\mp})^{\theta(\pm \ell+p)+N} \non \\ 
& & (T^{\pm})^{\theta(\pm \ell - p) +N } (S^{\mp})^{\theta(\pm \ell - p)} 
\, , \qquad 
(T^{\pm})^{\theta(\pm \ell - p)} (S^{\mp})^{\theta(\pm \ell - p) + N}
\label{op2} 
\eea
Here we assume that $\ell$ and $p$ are such integers or half-integers that  
$\ell \pm p$ are integers.

\par 
 In the sector  $S^Z \equiv \ell$ (mod $N$),  the operators 
 (\ref{op1}) and (\ref{op2}) commute with the twisted XXZ Hamiltonian
${\cal H}_{XXZ}(\phi)$  with $\phi \equiv p$ (mod $N$). 
We recall that ${\cal H}_{XXZ}(\phi)$ is given by 
the logarithmic derivative of the twisted transfer matrix $\tau(u; \phi)$.   
 For instance, 
we can show that $S^{\pm (N)}$  and $T^{\pm (N)}$ commute 
with the anti-periodic XXZ Hamiltonian 
in the sector  $S^Z \equiv N/2$ (mod $N$), 
 when $q$ is a primitive $2N$th root of unity.  Putting 
 $\ell = N/2$ and $p= N/2$ in (\ref{op2}),  
 we have 
\begin{eqnarray}
{[}S^{\pm(N)},H_{XXZ}(N/2) {]} \,  | N/2 \ra 
= {[}T^{\pm(N)},H_{XXZ}(N/2) {]} \, |N/2 \ra = 0 .
\end{eqnarray}

\subsection{Symmetry operators for 
the  inhomogeneous twisted transfer matrix } 

 From the equations (\ref{inhomo-main}), 
in the sector  $S^Z \equiv \ell ~ ({\rm mod} N)$, we have the commutation 
relations $X \tau(u; p; \{ w_j \} ) = 
\tau(u; p; \{ w_j \}) X$ where $X$s are given by     
\be 
({\tilde S}^{\pm})^{\theta(\pm \ell+p)} 
({\tilde T}^{\mp})^{\theta(\pm \ell+p)} \,,  \quad 
({\tilde T}^{\pm})^{\theta(\pm \ell - p)} 
({\tilde S}^{\mp})^{\theta(\pm \ell - p)}
\label{op1-inhomo}
\ee
and commutation or anti-commutation  relations $X \tau(u; p; \{ w_j \} ) 
= q^{N} \tau(u; p; \{ w_j \}) X$ where $X$s are given by 
\bea 
& & ({\tilde S}^{\pm})^{\theta(\pm \ell+p)+N} 
({\tilde T}^{\mp})^{\theta(\pm \ell+p)} \, , 
\qquad 
({\tilde S}^{\pm})^{\theta(\pm \ell+p)} 
({\tilde T}^{\mp})^{\theta(\pm \ell+p)+N} \non \\ 
& & ({\tilde T}^{\pm})^{\theta(\pm \ell - p) +N } 
({\tilde S}^{\mp})^{\theta(\pm \ell - p)} 
\, , \qquad 
({\tilde T}^{\pm})^{\theta(\pm \ell - p)} 
({\tilde S}^{\mp})^{\theta(\pm \ell - p) + N}
\label{op2-inhomo}
\eea
Here we recall that $\ell$ and $p$ are such integers or half-integers that  
$\ell \pm p$ are integers.

\subsection{The $sl_2$ loop algebra at $\Phi= \pi$}   

Let us discuss the anti-periodic boundary conditions 
or the twisted boundary conditions with $\Phi= \pi$. 
In the sector $S^Z \equiv N/2$ (mod $N$), 
we can show explicitly 
the defining relations of the $sl_2$ loop algebra.  
We consider  two cases: (i) $q$ is a primitive $2N$th root of 
unity with $N$ even ($L$ is even); 
(ii) $q$ is a primitive $2N$th root of 
unity with $N$ odd ($L$ is odd).

\par 
Let us consider the formula for generic $q$ 
\be 
{[} S^{+(N)}, S^{-(N)} {]} = \sum_{j=1}^{N} 
{\frac {S^{-(N-j)} S^{+(N-j)}} {[j]!}} \prod_{k=0}^{j-1} 
{\frac {q^{2S^Z -k} - q^{-2 S^Z +k}} {q-q^{-1}}} 
\ee
Taking the limit $q^{2N} \rightarrow 1$
in  the sector $S^Z \equiv N/2 ~ ({\rm mod}~N)$,  
we have the following  for the cases (i) and (ii):  
\begin{equation}
[S^{+(N)},S^{-(N)}]=[T^{+(N)},T^{-(N)}]= {2 \over N} S^Z 
\end{equation}
We thus take the following identification for the cases (i) and (ii):   
\begin{equation}
E_0^{+} = T^{-(N)}, \, E_0^{-}=T^{+(N)}, \, E_1^{+} = S^{+(N)}, \, 
E_1^{-} = S^{-(N)} , \, -H_0=H_1=  {\frac 2 N}  S^Z .    
\label{id2twist}
\end{equation}
Then, we can show that they satisfy the defining relations 
of the $sl_2$ loop algebra:  
(\ref{Cartan}), (\ref{EF}) and (\ref{Serre}).

\subsection{The $sl_2$ loop algebra symmetry of 
the periodic XXZ spin chain with $L$ odd}

As an application of the (anti-)commutation relations (\ref{STtr}),
 we discuss the case of the periodic boundary conditions 
with $L$ odd, where $q$ is a primitive $N$th root of unity with $N$ odd.   
Taking the same identification (\ref{id2twist}) of the generators,   
we can show explicitly that they satisfy 
the defining relations of the $sl_2$ loop algebra 
in the sector $S^Z \equiv N/2$ (mod $N$).  

\par 
 For sectors other than $S^Z \equiv N/2$ (mod $N$), we have not
 explicitly shown 
 the defining relations of the $sl_2$  loop algebra. 
However, we have a conjecture that some of the operators given in 
 (\ref{op2}) should generate the $sl_2$ loop algebra. 
By a different method, we can show 
that the spectral degeneracies related to the 
$sl_2$ loop algebra  also exist in the sectors 
other than $S^Z \equiv N/2$ (mod $N$). 
We can derive it from the general result on the spectral degeneracy 
of the eight-vertex model \cite{Missing} 
 and through the XXZ limit of the XYZ spin chain.  
Some details should be discussed elsewhere. 

{\vskip 1.2cm}
\par \noindent 
{\bf Acknowledgements }

The author would like to thank A.A. Belavin, C. Korff, K. Kudo, B.M. McCoy, 
and A. Nishino for their helpful comments.  
He would also like to thank the organizers: D. Arnaudon, J. Avan, 
L. Frappat, E. Ragoucy and P. Sorba 
for their kind invitation to the international workshop 
on Recent Advances in the Theory of Quantum Integrable Systems 
(RAQIS03), 25--28 March 2003, LAPTH, Annecy-le-Vieux, France. 
This work is partially supported by the Grant-in-Aid 
(No. 14702012).

\appendix
\section{Commutation or anti-commuation relations}

We can show the following relations for $q$ generic: 
\bea 
%
%
\Pi(\phi) \left( S^{\pm} \right)^n \Pi(\phi)^{-1} 
& = & \Big\{ (S^{\pm})^n + q^{\mp(n-1)}[n] (S^{\pm})^{n-1}S_1^{\pm} 
 \left( q^{2(S^Z \pm \phi)} -1 \right) \Big\} q^{-n\sigma_1^Z}  \non \\ 
%
%
 \Pi(\phi)^{-1} \left( S^{\pm} \right)^n \Pi(\phi)  
& = & \Big\{ (S^{\pm})^n + q^{\pm (n-1)}[n] (S^{\pm})^{n-1}S_L^{\pm} 
\left( q^{- 2(S^Z \pm \phi)} -1 \right) \Big\} q^{n \sigma_L^Z}  \non \\   
%
%
\Pi(\phi) \left( T^{\pm} \right)^n  \Pi(\phi)^{-1} 
& = & \Big\{ (T^{\pm})^n + q^{\pm(n-1)}[n] (T^{\pm})^{n-1} T_1^{\pm} 
\left( q^{-2(S^Z \mp \phi)} -1 \right) \Big\} q^{n \sigma_1^Z}  \non \\   
%
%
\Pi(\phi)^{-1} \left( T^{\pm} \right)^n \Pi(\phi)  
& = & \Big\{ (T^{\pm})^n + q^{\mp (n-1)}[n] (T^{\pm})^{n-1} T_L^{\pm} 
\left( q^{ 2(S^Z \mp \phi)} -1 \right) \Big\} q^{- n \sigma_L^Z}  \non
\\    
\label{PPP2}
\eea

 By using (\ref{PPP2}) we have  
\bea
& & (S^{\pm})^n \, \tau(u; \phi)  =  \tau(u; \phi + n) \, (S^{\pm})^n
\non \\
& & 
\qquad - q^{\mp (n-1)} [n]  (S^{\pm})^{n-1} S_1^{\pm} 
\left(q^{2(S^Z \pm n \pm \phi)} -1 \right) \, \Pi(\phi) X_{LL}^{+}(u)        
\non \\ 
& & \qquad - q^{\pm (n-1)}[n] X_{RR}^{+}(u) \,
 (S^{\pm})^{n-1} S_L^{\pm} 
\left(q^{- 2(S^Z \pm n \pm  \phi)} -1 \right) \,  \Pi(\phi)^{-1}        
\non \\
& & (T^{\pm})^n \, \tau(u; \phi) =  \tau(u; \phi-n) \, (T^{\pm})^n \non \\
& & \qquad - q^{\pm (n-1)}[n]  (T^{\pm})^{n-1} T_1^{\pm} 
\left(q^{- 2(S^z \pm n  \mp  \phi)} -1 \right) \, \Pi(\phi) X_{LL}^{-}(u)      
\non \\
& & \qquad - q^{\mp (n-1)}[n] X_{RR}^{-}(u) \,
(T^{\pm})^{n-1} T_1^{\pm} 
\left(q^{2(S^Z \pm n \mp \phi)} -1 \right) \, \Pi(\phi)^{-1}      
\label{STtr2}
\eea

\end{document}